\documentclass[conference]{IEEEtran}
\IEEEoverridecommandlockouts
\usepackage{cite}
\usepackage{amsmath,amssymb,amsfonts}
\usepackage{algorithmic}
\usepackage{graphicx}
\usepackage{textcomp}
\usepackage{xcolor}
\def\BibTeX{{\rm B\kern-.05em{\sc i\kern-.025em b}\kern-.08em
    T\kern-.1667em\lower.7ex\hbox{E}\kern-.125emX}}

\begin{document}
\title{Measurement of 2x2 LoS Terahertz MIMO Channel\\
\thanks{This work was funded by the NSF under contract CNS-1618936 and
  CNS-1910655.}}

\author{\IEEEauthorblockN{Suresh Singh}
\IEEEauthorblockA{Department of Computer Science\\
Portland State University\\
Portland, Oregon 97207\\
Email: singh@cs.pdx.edu}
\and
\IEEEauthorblockN{Thanh Le}
\IEEEauthorblockA{Department of ECE\\
Portland State University\\
Portland, Oregon 97207\\
Email: thanh4@pdx.edu}
\and
\IEEEauthorblockN{Ha Tran}
\IEEEauthorblockA{Department of ECE\\
Portland State University\\
Portland, Oregon 97207\\
Email: tranha@pdx.edu}
}

% make the title area
\maketitle

\begin{abstract}
This paper examines the performance of a 2x2 Line of Sight (LoS)
Multiple Input Multiple Output (MIMO) channel at three terahertz
frequencies -- 340 GHz, 410 Ghz, and 460 GHz. While theoretical models
predict very high channel capacities, we observe lower capacity which
is explained by asymmetric transmit-to-receive signal strengths as
well as due to signal attenuation over longer distances. Overall,
however, we note that at 460 Ghz, channel capacity of higher than 12
bps/hz is possible even at sub-optimal inter-antenna spacings (for
different distances). An important observation is also that we need to
maintain appropriate receive signal levels at receive antennas in
order to improve capacity.
\end{abstract}


\section{Introduction}

There is an increasing number of applications that would benefit from
very high data rate short-range wireless channels. Examples include
immersive virtual reality, backbone links in portable data centers,
backing up massive amounts of image data in real-time to the cloud,
etc. Indeed, many of these applications also require privacy from
snooping. Given the large available bandwidth from 0.1 to 10
terahertz and poor propagation properties due to absorption, the
terahertz spectrum appears to be ideally suited to meet these two
needs. Unfortunately, the poor propagation properties (including the
almost total lack of reflected paths) also means that
delivering high data rates will require highly directional
channels. This paper presents measurements of a 2x2 MIMO (Multiple
Input Multiple Output) channel for terahertz. The channel is a line of
sight (LoS) channel due to lack of refelcted paths in the collected
data.

Creating a high capacity LoS MIMO channel is possible if we can ensure
the orthogonality of the line of sight propagation paths
\cite{sarris05icicsp,jiang03vtc}. The key requirement is maintaining a
specific relationship between the frequency, antenna spacing and
distance between the communicating devices (section
\ref{model}). This unfortunately means that it is possible to
reduce the channel capacity by simple changes in transmitter receiver
distance or orientation. At low frequencies, this would require fairly
large scale changes in physical orientation and thus the channel is
fairly robust. However, at terahertz frequencies, where the wavelength
is of the order of millimeters, the channel becomes extremely
sensitive to any change in relative spatial orientation. In this paper
we examine this sensitivity of channel capacity to changes in
transmitter receiver distances using measurements.

The remainder of the paper is organized as follows. In the next
section we describe our system model and the measurement model that
was utilized. The following section (section \ref{setup}) describes the measurement setup
and we provide an analysis of the collected data in section \ref{results}. We also examine the
variation in capacity as a function of distances. Related work is
presented in section \ref{related} and conclusions in section
\ref{conclude}.

\section{MIMO LoS Model} \label{model}

Early work on LoS MOMO includes \cite{gesbert02tc,sarris05icicsp}
among others. Much of the work in the past has looked at LoS MIMO for
fixed backhaul microwave links and more recently for mimmimeter wave
frequencies at 60 GHz \cite{sheldon08ecwt,yu02iurs} as well. This and
other previous work provides us with a simple model for calculating
the capacity of LoS MIMO channels. Let us assume that both the sender and
receiver have $N$ antennas. Let the distance
between the $i$th transmitter antenna and the $j$th receiver antenna
be denoted as $d_{ij}$. The complex channel response between these two
elements can be written as $e^{-jkd_{ij}}/d_{ij}$, where
$k=2\pi/\lambda$. The complete channel response matrix can then be
written as \cite{sarris05icicsp},
\begin{equation}
{\cal H} = \left(
\begin{array}{llll}
e^{-jkd_{11}} &e^{-jkd_{12}} & \cdots &e^{-jkd_{1N}} \\
e^{-jkd_{21}} &e^{-jkd_{22}} & \cdots &e^{-jkd_{2N}} \\
. & . & \cdots & .\\
. & . & \cdots & .\\
. & . & \cdots & .\\
e^{-jkd_{N1}} &e^{-jkd_{N2}} & \cdots &e^{-jkd_{NN}} \\
\end{array}
\right)
\label{hmatrix}
\end{equation}
The channel capacity of the $N\times N$ LoS MIMO channel can then be
written as,
\begin{equation}
C = \log_2 \left(\det\left( {\cal I}_N + \frac{\rho}{N}{\cal H}{\cal
      H}^\dagger\right)\right)\mbox{ bps/Hz}
\label{capacity}
\end{equation}
where ${\cal I}_N$ is the $N\times N$ identity and $\rho$ is the SNR. The
capacity is maximized when ${\cal H}{\cal H}^\dagger = N{\cal I}_N$,
which corresponds to the case when all the individual sub-channels are
orthogonal and the system essentially behaves as $N$ SISO channels.

\subsection{2X2 LoS MIMO for Terahertz}

If we consider just a 2x2 system, the value of ${\cal
  H}{\cal H}^\dagger$ can be written as,
\begin{equation}
\left[
\begin{array}{cc}
2 & e^{jk(d_{21}-d_{11})}+e^{jk(d_{22}-d_{12})}\\
e^{jk(d_{11}-d_{21})}+e^{jk(d_{12}-d_{22})} & 2\\
\end{array}
\right]
\label{model2x2}
\end{equation}
If we assume that the sender and receiver arrays are parallel and the
inter-elemnent spacing is the same, then $d_{11}=d_{22}$ and
$d_{12}=d_{21}$. To obtain ${\cal
  H}{\cal H}^\dagger = 2{\cal I}_2$ we therefore need,
\[
|d_{11}-d_{12}| = (2p+1)\frac{\lambda}{4}
\]
where $p=0,1,2,\cdots$. If $s$ is the inter-antenna spacing and $d$ is
the distance between the two arrays, we obtain the following
expression that relates $s, d, \lambda$ for maximizing capacity,
\[
s^2 = \frac{d\lambda}{2}
\]

\begin{figure}[th]
\centerline
{\includegraphics[width=3.3in]{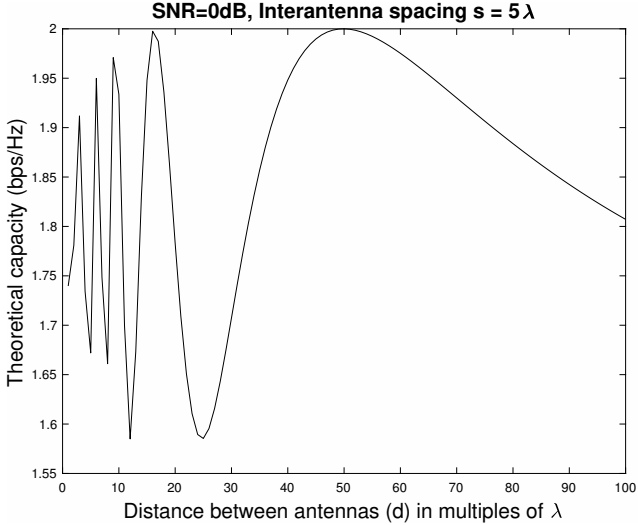}}
\caption{Theoretical capacity as a function of distance.}
\label{c2x2}
\end{figure}

In this paper we are interested in studying the performance os 2x2
MIMO when the antenna spacing is fixed (as it would in any actual
system) but the distance between the transmitter and receiver
varies. Let us use the above expression for capacity and set the SNR
$\rho = 0$dB and inter-antenna spacing at $5 \lambda$. Figure
\ref{c2x2} plots the maximum capacity obtainable as a function of
transmitter receiver separation. As the figure shows, the capacity
fluctuates very rapidly with increasing distance. Consider a frequency
of 410 Ghz (one of the frequency windows in the terahertz band that is
being considered for communications). The wavelength is 0.73mm which
means that the capacity will fluctuate as the transmitter receiver
distance varies by even small distances of the order of a few cm. In
the figure we also see that there are multiple distances at which
maximum capacity is achieved. These correspond to different $p$ values
above.

\begin{figure}[th]
\centerline
{\includegraphics[width=0.85\columnwidth]{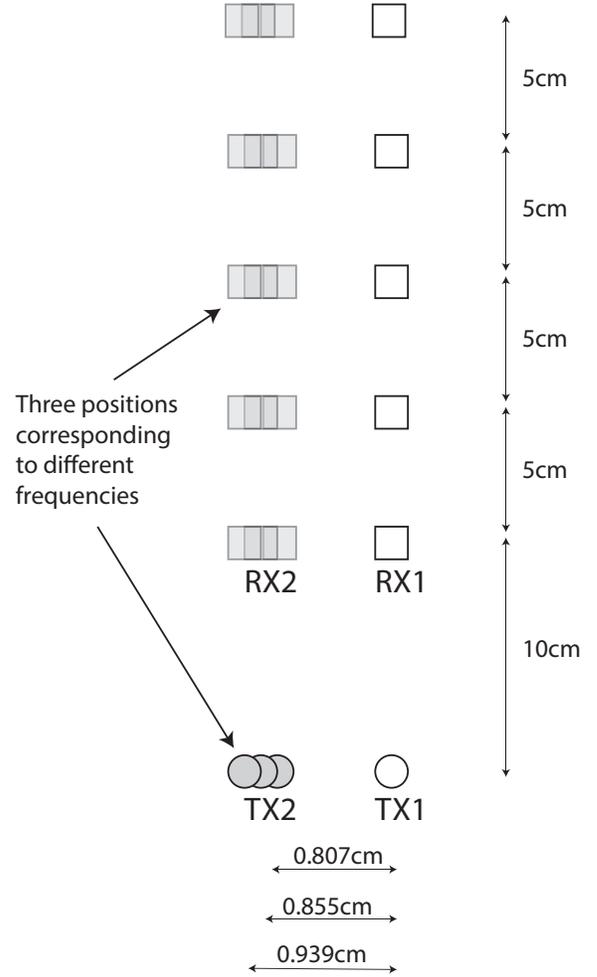}}
\caption{Experimental design.}
\label{design}
\end{figure}

In this paper we consider 2x2 MIMO systems where the antenna spacing
is fixed and the transmitter and receiver are at different
distances. We consider three different frequency bands that have been
identified as possible candidates for future communications
applications due to the fact that they do not suffer from molecular
absorption. The bands are 340 Ghz, 410 GHz, and 460 GHz. Figure
\ref{design} shows the experimental outline. TX1 indicates transmit
antenna 1 which is fixed while TX2 is the location of the second
transmit antenna. This second antenna is placed at three different
distances as shown. We computed the {\em optimal} distance between the
antennas based on the frequency and a distance of 20cm between the
transmitter and receiver. As the figure shows, we then measured the
received signal at the two receive antennas for five distances 10, 15,
20, 25, 30cm. 

\section{Measurement Setup} \label{setup}

\begin{figure}[th]
\centerline
{\includegraphics[width=3in]{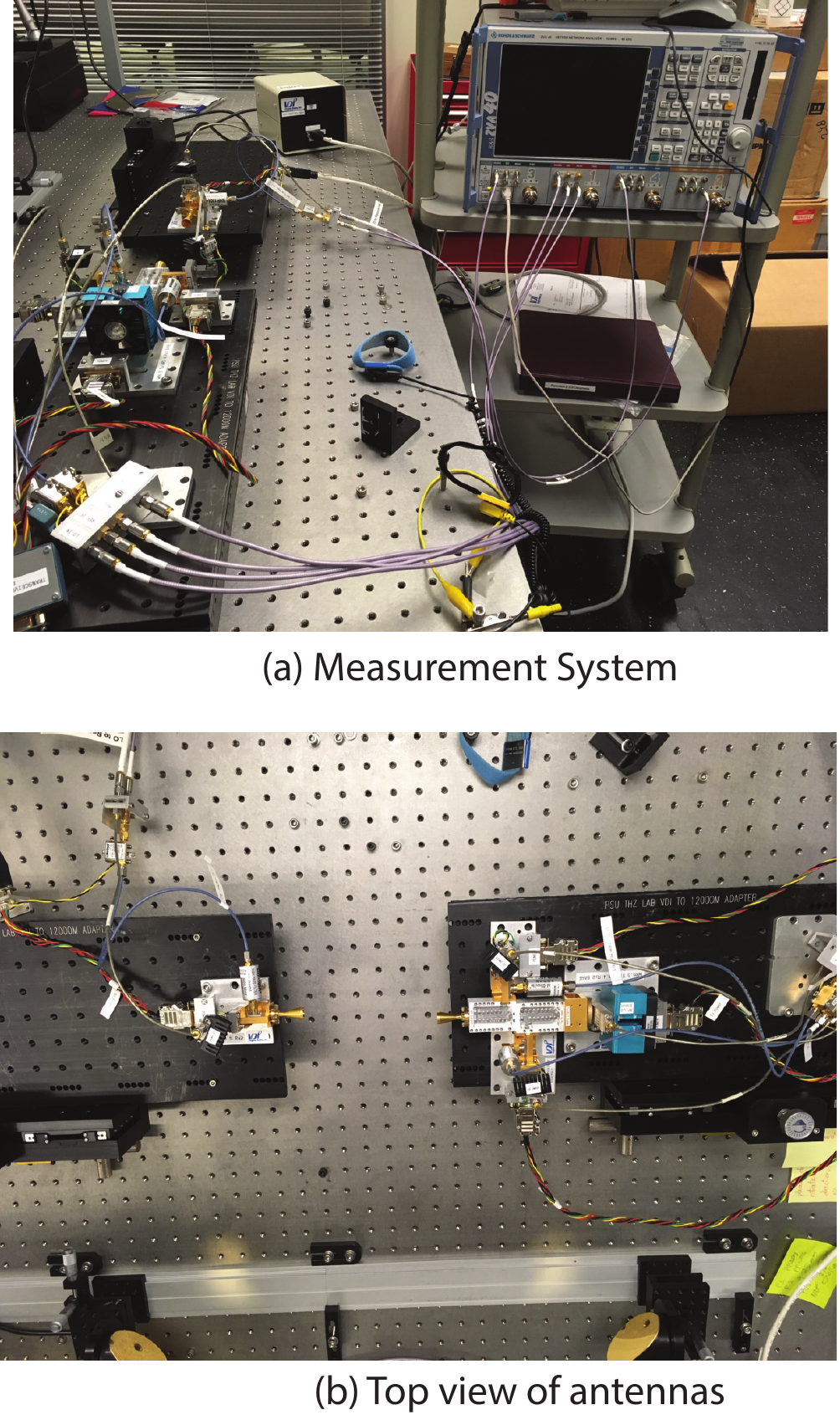}}
\caption{Measurement setup.}
\label{labsetup}
\end{figure}

We use a Rhode \& Schwartz Vector Network Analyzer (VNA) Figure
\ref{labsetup}(a). The setup is 
capable of producing signals up to 700 GHz. The transmit and receive
antennas we use are horn antennas as shown in Figure
\ref{labsetup}(b) with gain 25 dBi and a $12^0$ half beam width
(HBW). Our measurements were conducted in a environmentally controlled
lab setting where the temperature was held constant at 72F and the
humidity was 40\%. The other experimental parameters utilized were:
\medskip

\begin{tabular}{|l|l|} \hline
Output power & 5 dbm\\
Center Frequency & 340, 410, 460 Ghz\\
Inter-antenna distance& 0.939, 0.855, 0.807 cm\\
Tx-Rx distance used for &\\
inter-antenna separation& 20 cm\\
Number of points & 1751\\
IF Bandwidth & 1 kHz\\
Averaging & 10\\ \hline
\end{tabular}
\medskip

\begin{figure}[th]
\centerline
{\includegraphics[width=\columnwidth]{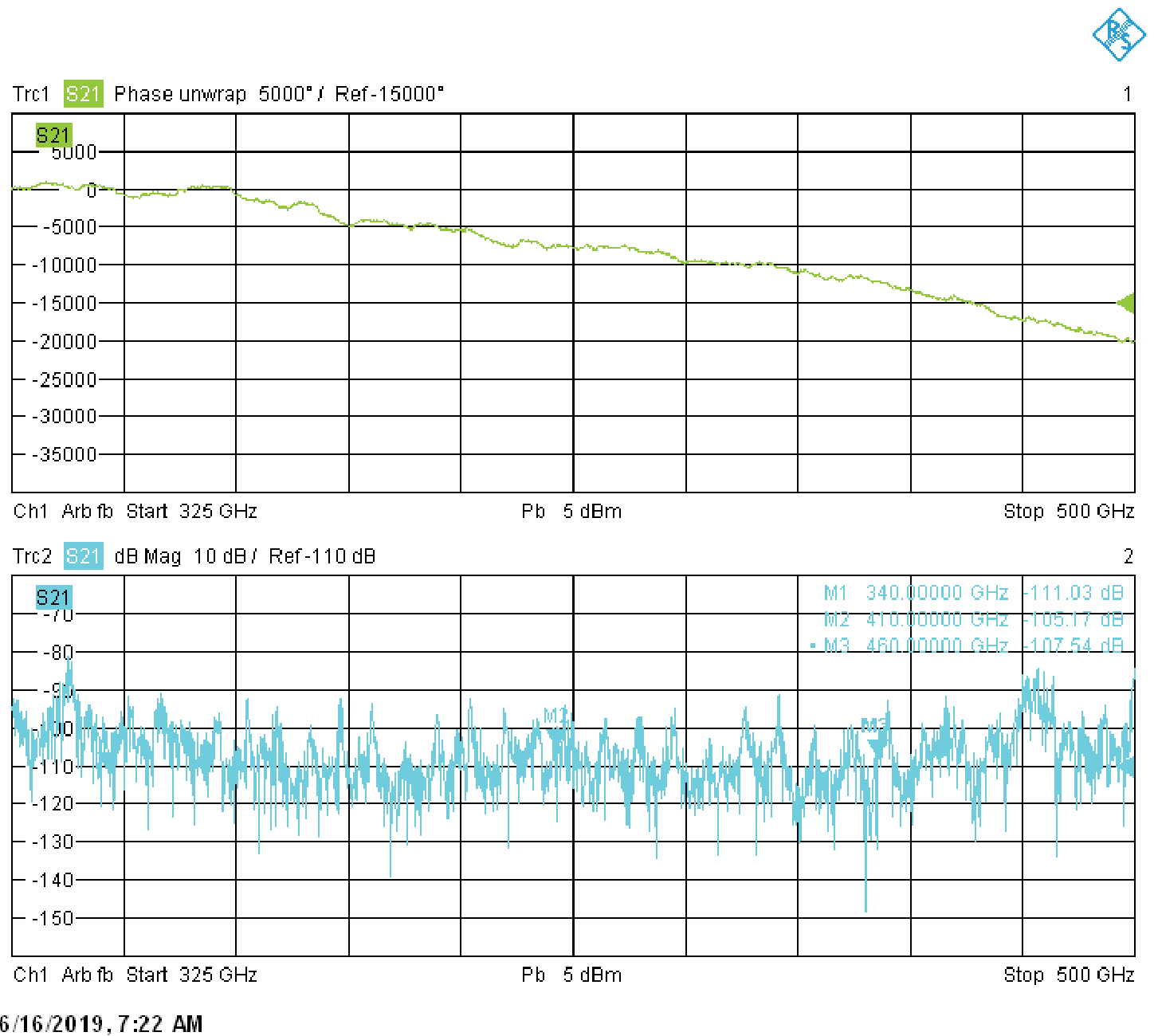}}
\caption{Noise floor.}
\label{noise}
\end{figure}

\begin{figure}[th]
\centerline
{\includegraphics[width=\columnwidth]{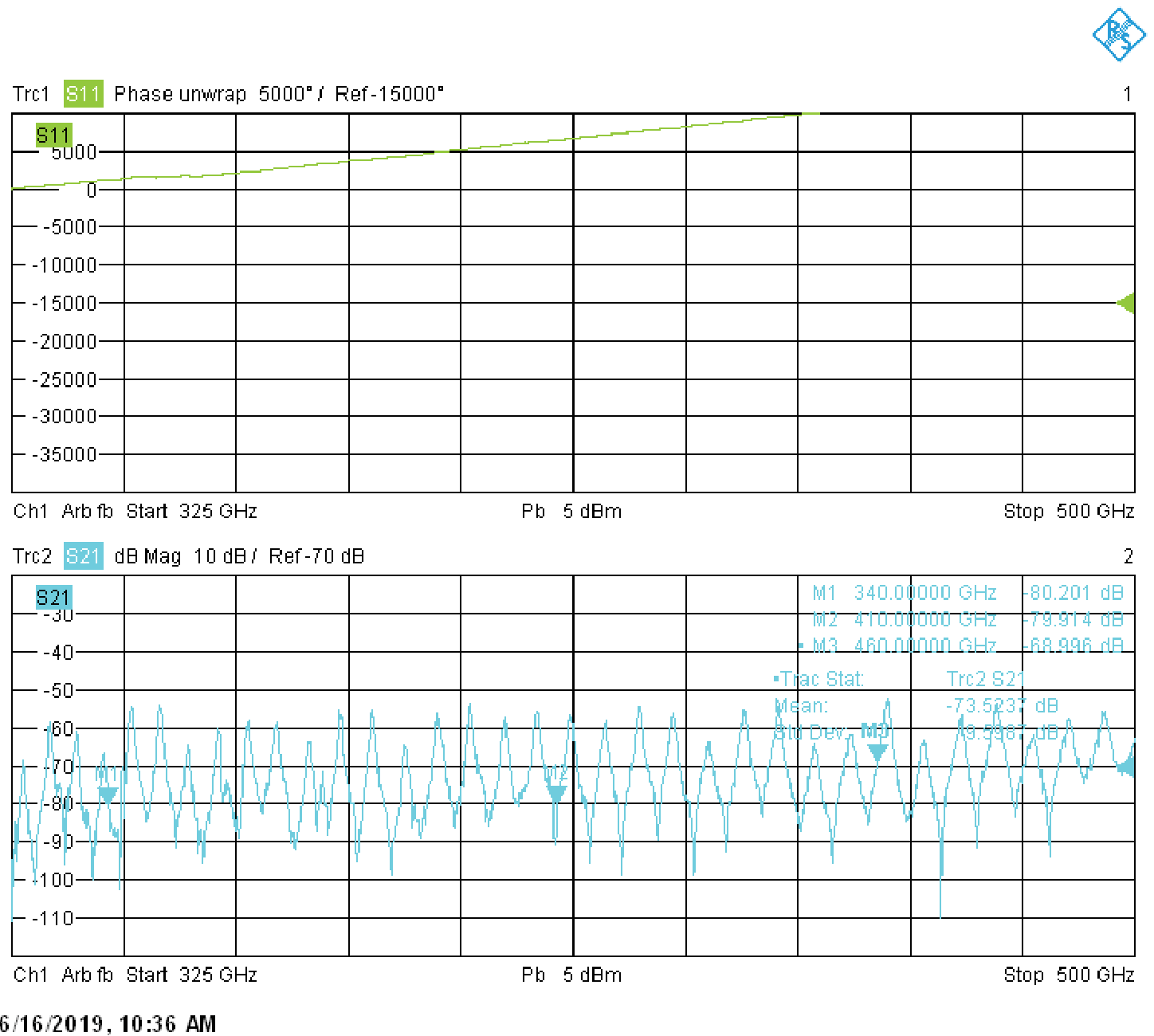}}
\caption{Signal level at a distance of 30cm.}
\label{signal}
\end{figure}

We measured the noise floor at each frequency by shorting Tx1. The
measured SNR at the maximum distance of 30 cm for the three
frequencies is 30.83 dB (340 GHz), 25.2 dB (410 GHz), and 38.54 dB
(460 GHz). Figure \ref{noise} plots the noise floor at 30 cm for 325 -
500 GHz while Figure \ref{signal} plots the signal level for the same
frequency range at 30 cm.

\section{Main Results} \label{results}

\begin{figure}[th]
\centerline
{\includegraphics[width=\columnwidth]{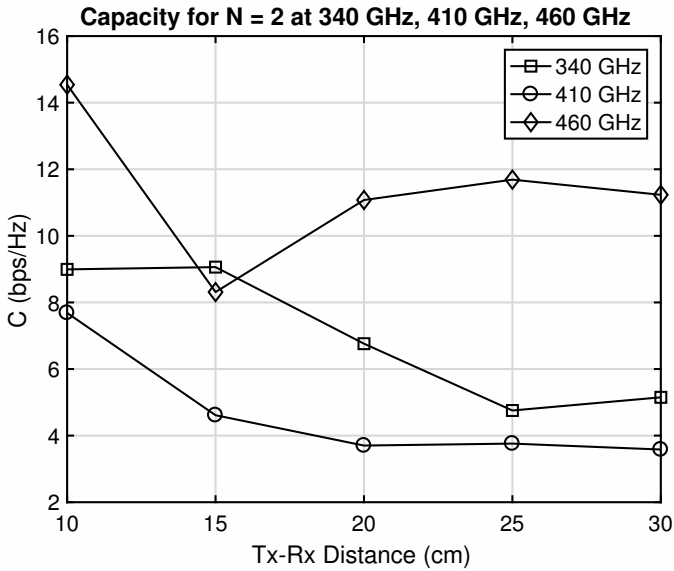}}
\caption{Estimated capacity using channel measurements.}
\label{measuredcapacity}
\end{figure}

From the collected data, we extracted the received signal phase and
amplitude for the three frequencies at the five distances and then
used equation \ref{capacity} to obtain the theoretical capacity in
bps/hz. The results are plotted in Figure \ref{measuredcapacity}. The
capacity of the three chennels is varies significantly with distance
though the 460 GHz channel capacity improves at 20-25cm. This is
explained when we observe from Figure \ref{c2x2} that there are
multiple distances at which a fixed inter-antenna spacing is
optimal. On the other hand, the question is why do we not see the same
behavior for the other frequencies and why is the maximum not at 20 cm
for which the inter-antenna spacing was optimized.

\begin{figure}[th]
\centerline
{\includegraphics[width=\columnwidth]{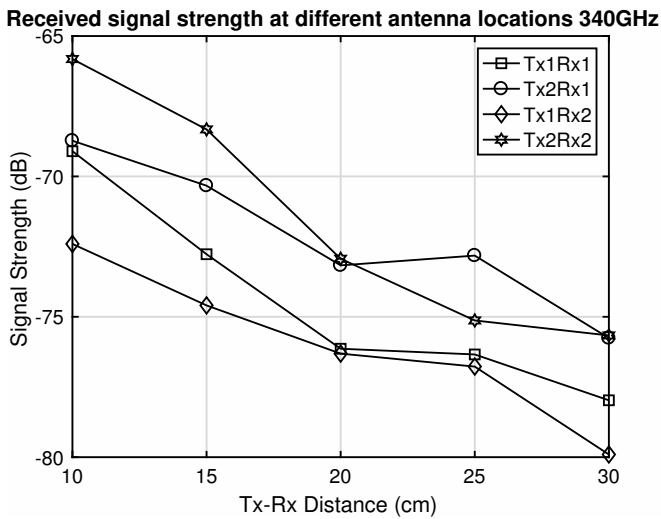}}
\caption{Signal strength for various antenna pairs (340GHz).}
\label{340signal}
\end{figure}

\begin{figure}[th]
\centerline
{\includegraphics[width=\columnwidth]{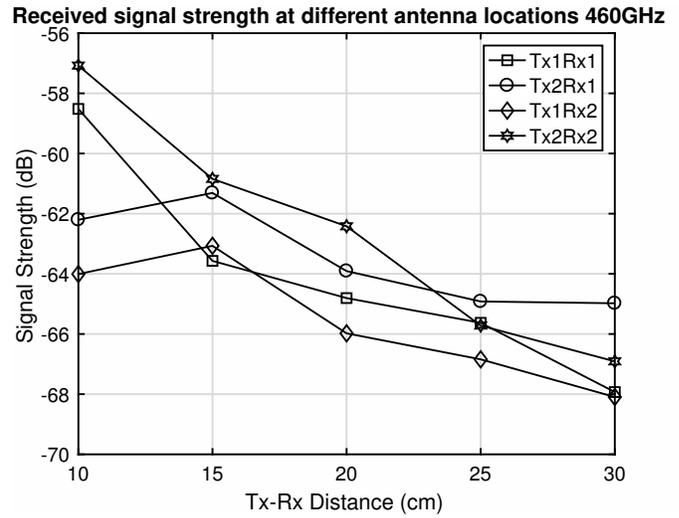}}
\caption{Signal strength for various antenna pairs (460GHz).}
\label{460signal}
\end{figure}

To explain these two discrepancies, it is helpful to plot the signal
strength at different distances and for different combinations of
antennas. In Figure \ref{340signal} we plot this data for the 340GHz
channel. First, note that the signal strength decreases with distance
while we assume a constant SNR in the optimal plot in Figure
\ref{c2x2}. The other thing of note is that the received signal
strength from transmit antenna 2 is greater than that from
transmit antenna 1. This means the ${\cal HH'}$ matrix cannot be
$2{\cal I}_2$ which reduces capacity as well as changes the
relationship between the optimal antenna spacing and transmit to
receive distances. Finally, we plot the signal strength for 460 Ghz in
Figure \ref{460signal} to compare against Figure \ref{340signal}. The
signal strength is much higher for the 460 Ghz band which explains its
higher data rate overall. However, as in the case of the other
frequencty bands, we see an asymmetrix receiver signal strength
behavior. 

\section{Related Work} \label{related}

For many years, the primary focus of terahertz research was on
material sensing. For instance, due to the specific molecular
absorption properties of various explosives, terahertz has been used
to detect trace amounts of different compounds that are used in their
manufacture. Terahertz has also been shown to be very effective in
imaging the skin as a way to detect skin cancer. Airports utilize
terahertz to image under clothing to search for weapons. Recently,
terahertz has been employed to measure the thickness of coatings on
medicines or automotive paint, etc \cite{Federici2005,Woolard2008}.

Research in utilizing terahertz for communications has lagged
primarily due to the difficulty in manufacturing communication systems
at these frequencies. The two commonly used devices for channel
characterization that we utilize, for instance, are very expensive and
are limited in power output. Recently, however, things have begun to
change. For instance, Fujitsu recently demonstrated a compact 20 Gbps
radio operating at 300 GHz \cite{fujitsu2017}. Other work, such as
\cite{merkle17csics} recently demonstrated a 4-element phased antenna
array operating from 275 - 325 GHz.

Most of the research on terahertz communications has focused on
point-to-point links utilizing simple modulation methods such as OOK
(On Off Keyeing) and PSK (Phase Shift Keyeing). For example, 10 GBps
data rates at 120 GHz were demonstrated in \cite{Hirata2010} for 
distances up to 6km. At the other end of the transmission range,
\cite{Kallfass2011} demonstrated transmissions using OOK at 220 GHz
and achieved 50 Gbps over 50 cm. \cite{Song2009} achieved 8 Gbps at
250 GHz using ASK modulation. Recently, \cite{jia18jlt} demonstrated
106 Gbps at 400 GHz using a photonic wireless link. There have been
other noteworthy demonstrations as well including
\cite{Antes2011,Zhang2012,Song2012,Ducournau2014,Wang2013,Lu2013}. Finally,
a significant milestone was achieved in \cite{Koenig2013} where they
achieved 100 Gbps over 20m using a phased antenna array.

Our work is among the few to consider MIMO for terahertz
frequencies. \cite{khalid16wcl} was one of the earliest to study MIMO
for 5G at 298-313 GHz. They utilized a measurment system similar to
ours to show that a MIMO system is possible. However, they did not
consider the impact of varying distances or other factors on the
achievable channel capacity.

\section{Conclusions} \label{conclude}

We present initial results of ongoing measurements for terahertz MIMO
systems. We presented results for LoS MIMO capacity for three
frequency bands in the terahertz spectrum that are being considered
as possible communication windows. As expected, we note that
increasing distance between transmitter and receiver results in
lowering of capacity for all three frequency bands though 460 GHz
appears to suffer the least degradation (14 bps/hz at 10 cm to 12
bps/hz at 25 cm). We observed that the receiver signal strength is
asymmetric (i.e., Tx1 to Rx1 is weaker than Tx2 to Rx2, for
example). This is something we are investigating currently as it
results in unexpected capacity behavior. We are also extending our
measurements to other frequency windows as well as to larger antenna
arrays.

%%%%%%%%% references section%%%%%%
\bibliographystyle{abbrv}
\bibliography{reference,5g,thz,mynsfpubs}

\begin{thebibliography}{10}

\bibitem{Antes2011}
J.~Antes, J.~Reichart, D.~Lopez-Diaz, A.~Tessmann, F.~Poprawa, F.~Kurz,
  T.~Schneider, H.~Massler, and I.~Kallfass.
\newblock System concept and implementation of a {mmW} wireless link providing
  data rates up to 25 {Gbit/s}.
\newblock In {\em 2011 IEEE International Conference on Microwaves,
  Communications, Antennas and Electronics Systems (COMCAS)}, pages 1--4, Nov
  2011.

\bibitem{Ducournau2014}
G.~Ducournau, P.~Szriftgiser, A.~Beck, D.~Bacquet, F.~Pavanello, E.~Peytavit,
  M.~Zaknoune, T.~Akalin, and J.-F. Lampin.
\newblock Ultrawide-bandwidth single-channel 0.4-{THz} wireless link combining
  broadband quasi-optic photomixer and coherent detection.
\newblock 2014.

\bibitem{merkle17csics}
T.~M. et~al.
\newblock Testbed for phased array communications from 275 to 325 ghz.
\newblock In {\em Proceedings IEEE Compound Semiconductor Integrated Circuit
  Symposium (CSCIS)}, 22-25 October 2017.

\bibitem{Federici2005}
J.~F. Federici, D.~Gary, R.~Barat, and D.~Zimdars.
\newblock {THz} standoff detection and imaging of explosives and weapons.
\newblock In {\em Defense and Security}, pages 75--84. International Society
  for Optics and Photonics, 2005.

\bibitem{gesbert02tc}
D.~Gesbert, H.~Bolcskei, D.~Gore, and A.~Paulraj.
\newblock Outdoor mimo wireless channels: models and performance prediction.
\newblock {\em IEEE Transactions on Communications}, 50(12):1926--1934,
  December 2002.

\bibitem{Hirata2010}
A.~Hirata, T.~Kosugi, H.~Takahashi, J.~Takeuchi, K.~Murata, N.~Kukutsu,
  Y.~Kado, S.~Okabe, T.~Ikeda, F.~Suginosita, et~al.
\newblock 5.8-km 10-{Gbps} data transmission over a 120-{GHz}-band wireless
  link.
\newblock In {\em 2010 IEEE International Conference on Wireless Information
  Technology and Systems (ICWITS),}, pages 1--4. IEEE, 2010.

\bibitem{jia18jlt}
S.~Jia, X.~Pang, O.~Ozolins, X.~Yu, H.~Hu, J.~Yu, P.~Guan, F.~D. Ros, S.~Popov,
  G.~Jacobsen, M.~Galili, T.~Morioka, D.~Zibar, and L.~K. Oxenlowe.
\newblock 0.4 thz photonic-wireless link with 106 gb/s single channel bitrate.
\newblock {\em Journal of Lightwave Technology}, 36(2):610--616, January 2018.

\bibitem{jiang03vtc}
J.-S. Jiang and M.~A. Ingram.
\newblock Distributed source model for short-range mimo.
\newblock In {\em Proceedings IEEE Vehicular Technology Conference}, October
  2003.

\bibitem{Kallfass2011}
I.~Kallfass, J.~Antes, T.~Schneider, F.~Kurz, D.~Lopez-Diaz, S.~Diebold,
  H.~Massler, A.~Leuther, and A.~Tessmann.
\newblock All active {MMIC}-based wireless communication at 220 {GHz}.
\newblock {\em IEEE Transactions on Terahertz Science and Technology},
  1(2):477--487, November 2011.

\bibitem{khalid16wcl}
N.~Khalid and O.~B. Akan.
\newblock Experimental throughput analysis of low-thz mimo communication
  channel in 5g wireless networks.
\newblock {\em IEEE Wireless Communication Letters}, 5(6):616--619, December
  2016.

\bibitem{Koenig2013}
S.~Koenig, D.~Lopez-Diaz, J.~Antes, F.~Boes, R.~Henneberger, A.~Leuther,
  A.~Tessmann, R.~Schmogrow, D.~Hillerkuss, R.~Palmer, T.~Zwick, C.~Koos,
  W.~Freude, O.~Ambacher, J.~Leuthold, and I.~Kallfass.
\newblock Wireless sub-{THz} communication system with high data rate.
\newblock {\em Nature Photonics}, 7:977--981, October 2013.

\bibitem{Lu2013}
B.~Lu, W.~Huang, C.~Lin, and C.~Wang.
\newblock A {16QAM} modulation based {3Gbps} wireless communication
  demonstration system at 0.34 {THz} band.
\newblock In {\em Infrared, Millimeter, and Terahertz Waves (IRMMW-THz), 2013
  38th International Conference on}, pages 1--2. IEEE, 2013.

\bibitem{fujitsu2017}
Y.~Nakasha, S.~Shiba, Y.~Kawano, and T.~Takahashi.
\newblock Compact terahertz receiver for short-range wireless communications of
  tens of gbps.
\newblock {\em Fujitsu Science and Technology Journal}, 53(2):9--14, Feburary
  2017.

\bibitem{sarris05icicsp}
I.~Sarris and A.~R. Nix.
\newblock Maximum mimo capacity in line-of-sight.
\newblock In {\em International Conference on Information Communications and
  Signal Processing}, Bankok, Thailand, 6-9 December 2005.

\bibitem{sheldon08ecwt}
C.~Sheldon, E.~Torkildson, M.~Seo, C.~P. Yue, M.~Rodwell, and U.~Madhow.
\newblock Spatial multiplexing over a line-of-sight millimeter-wave mimo link:
  two-channel hardware demonstration at 1.2gbps over 41m range.
\newblock In {\em European Conference on Wireless Technology}, pages 198--201,
  27-28 October 2008.

\bibitem{Song2009}
H.~Song, K.~Ajito, A.~Hirata, A.~Wakatsuki, T.~Furuta, N.~Kukutsu, and
  T.~Nagatsuma.
\newblock Multi-gigabit wireless data transmission at over 200-{GHz}.
\newblock In {\em 34th International Conference on Infrared, Millimeter, and
  Terahertz Waves, 2009. IRMMW-THz 2009}, pages 1--2, 2009.

\bibitem{Song2012}
H.~Song, K.~Ajito, Y.~Muramoto, and A.~Wakatasuki.
\newblock 24 {Gbit/s} data transmission in 300 {GHz} band for future terahertz
  communications.
\newblock {\em Electronics Letters}, 48(15):953--954, July 2012.

\bibitem{Wang2013}
C.~Wang, C.~Lin, Q.~Chen, B.~Lu, X.~Deng, and J.~Zhang.
\newblock A 10-{Gbit/s} wireless communication link using 16-{QAM} modulation
  in 140-{GHz} band.
\newblock {\em {IEEE} Transactions on Microwave Theory and Techniques},
  61(7):2737--2746, July 2013.

\bibitem{Woolard2008}
D.~Woolard, P.~Zhao, C.~Rutherglen, Z.~Yu, P.~Burke, S.~Brueck, and A.~Stintz.
\newblock Nanoscale imaging technology for {THz}-frequency transmission
  microscopy.
\newblock {\em International Journal of High Speed Electronics and Systems},
  18(01):205--222, 2008.

\bibitem{yu02iurs}
K.~Yu, M.~Bengtsson, B.~Ottersten, and M.~Beach.
\newblock Narrowband mimo channel modeling for los indoor scenarios.
\newblock In {\em Proceedings 27th Triennial General Assembly of the
  International Union of Radio Science}, Maastricht, The Netherlands, August
  2002.

\bibitem{Zhang2012}
B.~Zhang, Y.-Z. Xiong, L.~Wang, and S.~Hu.
\newblock A switch-based {ASK} modulator for 10 {Gbps} 135 {GHz} communication
  by 0.13 {MOSFET}.
\newblock {\em IEEE Microwave and Wireless Components Letters}, 22(8):415--417,
  2012.

\end{thebibliography}

\end{document}